\documentstyle[aps,twocolumn]{revtex}
\tighten
\begin{document}

\preprint{HUTP-03/A041,hep-th/0306108}
\title{A Dynamical Approach to the Cosmological Constant}
\author{Shinji Mukohyama and Lisa Randall}
\address{
Department of Physics, Harvard University\\
Cambridge, MA, 02138, USA
}
\date{\today}

\maketitle

\begin{abstract}
 We consider a dynamical approach to the cosmological constant.
 There is a scalar field with a potential whose minimum occurs at
 a generic, but negative, value for the vacuum energy, and it has a
 non-standard kinetic term whose coefficient diverges at zero
 curvature as well as the standard kinetic term. Because of the
 divergent coefficient of the kinetic term, the lowest energy state is
 never achieved. Instead, the cosmological constant automatically stalls
 at or near  zero. The merit of this model is that it is stable under
 radiative corrections and leads to stable dynamics, despite the
 singular kinetic term. The model is not complete, however, in that some
 reheating is required. Nonetheless, our approach can at the very least
 reduce fine-tuning  by $60$  orders of magnitude or provide a new
 mechanism for sampling possible cosmological constants and implementing
 the anthropic principle. 
 \hfill\mbox{[HUTP-03/A041]}
\end{abstract}

\vspace{0.5cm}


The cosmological constant remains the biggest puzzle plaguing
particle physics~\cite{reviews,Witten}. There is no known mechanism that
guarantees zero, or nearly zero energy in a stable or metastable minimum
energy configuration. Recently, the problem of the existence of small
vacuum energy, whose value has to be about the same size as the energy
in matter {\it today}, has further complicated the
issue~\cite{Perlmutter,Schmidt,Riess}. The apparent impossibility of 
addressing these problems has led to an increase in  speculation for the
necessity of the anthropic principle~\cite{Anthropic}. Before resorting
to this, it is worthwhile to ask whether anything could possibly do what
the cosmological constant data requires~\cite{Witten}. It seems likely
that the correct way to interpret the  tiny value of the cosmological
constant is that conventional quantum field theory is not the whole
story, so it is worth seeking acceptable modifications.

One might be tempted to consider  non-standard
potentials. However, those are never stable under radiative corrections.
In this letter, we consider a new approach to the cosmological constant
problem in which we try to avoid a fine-tuned  potential,
or one that would be unstable against radiative corrections. We propose
that the apparent value of the cosmological constant is determined by
dynamical considerations. The true value of the vacuum energy is 
{\it not} zero. But the dynamics is such that the true minimum is never
attained, and the universe would settle down to a near zero energy
state.

In this approach to the cosmological constant, we assume that inflation
has already occurred, but ended when the vacuum energy was still large
and positive. We propose scalar dynamics that decrease the cosmological
vacuum energy in such a way that the scalar field stalls when the
curvature becomes close to zero. The effective cosmological constant is 
therefore zero or slightly positive.

We can think of two reasons this might happen. One possibility is
that there is a scaling solution to scalar dynamics in which the
kinetic and potential energies of a scalar field decrease in
tandem, so that the field is slowed down when small vacuum energy
is achieved. In our specific realizations, there was always an
instability invalidating the models we tried.

The second possibility is that there is a feedback mechanism in which
the equations for the scalar field depend on curvature in such a way
that the field stalls at or near zero vacuum energy. This is the
possibility that will be considered here.


Clearly, ordinary scalar field dynamics does not act the way we
postulate. We assume the coefficient of a kinetic term diverges at zero
curvature, so that the field gets frozen when the curvature approaches
zero. Without such a singular term, the scalar field would  overshoot
and the energy would become negative. The singular kinetic term makes
the scalar field stop at zero curvature, even without a fine-tuned
potential. 

The stability under radiative corrections is an important feature of
our model. Radiative corrections will produce additional regular
terms in the action, but the field will stall whether or not
these are present.

One might also worry about additional singular potential terms being
generated through radiative corrections but this does not happen. This
can be seen explicitly using the Lagrangian below in  Feynman diagrams
or by a change of variables for which the kinetic term will be
nonsingular, analogous to the Mukhanov choice of variable
\cite{Mukhanov} in standard field theory cosmology.

The unconventional Lagrangian we consider is
%
\begin{eqnarray}
 I & = & \int d^4x\sqrt{-g}
  \left[ \frac{R}{2\kappa^2} + \alpha R^2 + L_{kin} - V(\phi) \right],
  \nonumber\\
 L_{kin} & = & \frac{\kappa^{-4}K^q}{2qf^{2q-1}},
  \quad  
  \label{eqn:action}
\end{eqnarray}
where $f$ is a function of the Ricci scalar $R$ which vanishes at $R=0$
and behaves near $R=0$ as 
%
\begin{equation}
 f(R) \sim (\kappa^4R^2)^m, \label{eqn:fnear0}
\end{equation}
$\kappa$ is the Planck length, $\alpha$ and $q$ are constants, and
$K \equiv -\kappa^4\partial^{\mu}\phi\partial_{\mu}\phi$. 
We do not give a reason why $f(R)$ vanishes at $R=0$. However, we shall
show below that this choice of $f(R)$ is radiatively stable, leads to a
stable dynamics and could help solve the cosmological constant
problem. Our sign convention for the metric is $(-+++)$. We could make
$f$ dependent of $\phi$, but such a dependence can be removed from the
behavior of $L_{kin}$ near $R=0$ by redefinition of $\phi$ without loss
of generality. Note that $L_{kin}$ above represents the term which is
the most singular-looking at $R=0$ among many possible terms in the
kinetic part and that we did not include less singular-looking terms
since they are not important at low energy. We also omitted all other
dynamical fields since, as we shall see below, the dynamics of $\phi$ is
so slow that any dynamical fields other than $\phi$ will settle into
their ground state before the universe approaches a sufficiently low
energy state.

The goal is for the scalar field to stop rolling at or near zero
vacuum energy without any fine-tuning of its potential. For this
purpose, the minimum of the potential $V(\phi)$ should be negative
so that $V(\phi)$ has a root. We can absorb any nonzero cosmological
term into $V(\phi)$ without loss of generality as far as the minimum
remains negative. We would like to stress again that all dynamical
fields other than $\phi$ already settled into their ground state before
the universe approaches a low energy state and that $V(\phi)$ includes
the ground state energies of all such fields.

Notice there is no tuning in the potential. Any potential will work, 
so long as it has a minimum at negative energy.  However, we do
require a special form for the kinetic term.  If all coefficients
in the kinetic part are regular at zero curvature then it is
evident that $\phi$ never stops there. Hence, we are forced to consider
a singular kinetic term in which
 $f(R)$ vanishes at $R=0$. The only alternative would  be to have $f$
depend on $\phi$ so that $f$ vanishes at the root of $V(\phi)$.
However, this would not be stable under the addition of an extra
vacuum energy to $V(\phi)$. Namely, $\phi$ does not know where to
stop, whereas the curvature does.

We do not know a parent theory that will provide our Lagrangian as the
low-energy effective theory. We treat this model as a purely
phenomenological suggestion that might motivate further research into
the possible parent theory, which is presumably not based entirely on
conventional four-dimensional field theory.


We now show that this Lagrangian gives a feedback mechanism that
makes the field stall at zero vacuum energy. We restrict this
discussion  to $q=1$. It is easy to see how the mechanism works
for more general $q$.

In the flat Friedmann-Robertson-Walker background
%
\begin{equation}
 ds^2 = -dt^2 + a(t)^2(dx^2+dy^2+dz^2),
\end{equation}
the equation of motion for a homogeneous $\phi$ is
%
\begin{equation}
 \dot{\pi} + 3H\pi + V'(\phi) = 0, \quad
 \pi \equiv \dot{\phi}/f,
\end{equation}
where $H=\dot{a}/a$, a dot denotes the time derivative, and $\pi$ is the
momentum conjugate to $\phi$. As we shall see soon, $\phi$ evolves very
slowly and, thus, $V$ can be approximated by a linear function near
$V=0$ as
%
\begin{equation}
 V \simeq c\kappa^{-3}(\phi-\phi_0),
\end{equation}
where $c$ ($=O(1)$) and $\phi_0$ are constants. Without any fine-tuning,
the dimensionless constant $c$ should be of order unity. Hence, the 
asymptotic behavior of $\pi$ is
%
\begin{equation}
 \kappa^2\pi \sim -c\kappa^{-1}H^{-1}.
  \label{eqn:asymptotic-pi}
\end{equation}
This follows when $\dot{H}/H^2\simeq const.<0$ because the
momentum $\pi$ under the influence of a constant force ($V'\simeq
-c\kappa^{-3}$) is asymptotically proportional to time and, thus,
to $H^{-1}$. When $\dot{H}/H^2\simeq 0$, the above asymptotic form
for $\pi$ is a consequence of  the friction force $-3H\pi$, which
cancels the constant force when $3H\pi+c\kappa^{-3}=0$. We shall
soon see that $\dot{H}/H^2\simeq 0$ at low energy.
If the kinetic term is small compared to the potential
term then the Friedmann equation implies that
%
\begin{equation}
 3H^2 \simeq \kappa^2 V, \label{eqn:Friedmann-eq}
\end{equation}
and (\ref{eqn:asymptotic-pi}) can be rewritten as
%
\[
  \kappa \partial_t(\kappa^4V) \sim
  c\kappa^2\pi f \sim -c^2(\kappa H)^{4m-1} \sim -c^2(\kappa^4V)^{2m-1/2}.
\]
Hence, we obtain
%
\begin{equation}
 (\kappa^4V)^{-2m+3/2} \sim
  c^2\frac{t-t_0}{\kappa},
\end{equation}
where $t_0$ is a constant. If  $-2m+3/2<0$ then we achieve
%
\begin{equation}
 \kappa^4V   \to +0 \quad (t/\kappa\to\infty).
  \label{eqn:Vto0}
\end{equation}
This means that $\phi$ stalls at $V=0$, where $V$ includes all
contributions to the cosmological constant. Notice that the result
(\ref{eqn:Vto0}) is independent of the values of $c$ and $\phi_0$. We
also see that for a large $m$ ($-2m+3/2<-1/2$), $V$ approaches zero more
slowly than $t^{-2}$ so that $\dot{H}/H^2\simeq 0$ at low energy, which
was used above. We shall see later in this paper that stability requires
(for $q=1$)
%
\begin{equation}
 m > 3/2.
  \label{eqn:stabilityq=1}
\end{equation}
With this condition, it is also easy to see that $V$ does not jump to a
negative value by quantum fluctuation~\cite{paper2}.

We have assumed that the kinetic energy is small compared to the
potential energy. This assumption is easily justified. At low
energy ($\kappa H\ll 1$), if $m>1$ then 
%
\begin{eqnarray}
 L_{kin} & = &
  \frac{f}{2}\pi^2
  \sim \frac{c^2}{\kappa^4}(\kappa H)^{4m-2}
 \ll \frac{H^2}{\kappa^2} \sim V.
   \label{eqn:Lkin-is-small}
\end{eqnarray}


We have achieved the vanishing cosmological constant in a way
that is stable under radiative corrections and that has
self-consistent, stable dynamics. However, although the cosmological
constant approaches zero, it does so more slowly than matter or
radiation so that without additional structure, the universe would be
empty. It is not entirely clear that a dynamical model where this is not
the case could be successful since it is this property that makes it
possible for all fields other than $\phi$ to settle into their ground
state before $\phi$ stalls at zero curvature so that the zero curvature
really corresponds to the vanishing cosmological constant. Moreover, the
large $m$ and, thus, the slow evolution of $\phi$ are required for
stability. This does imply, however, that should this mechanism be
responsible for a low cosmological constant, reheating would
be required to thermally populate the universe after the cosmological
constant has decreased to a small value.
A couple of
possibilities for the reheat process are:

(I) Low-energy inflation. One can consider an extra scalar field $\chi$
with mass $m_{\chi}\sim 10^{-3}eV$ and a term like $-R\chi^2$. When 
$R\sim m_{\chi}^2$, a phase transition would occur (as in hybrid 
inflation \cite{Lindebook}) and the universe would be reheated up to
temperature $\sim TeV$. 
This phase transition happens when the energy stored in $\chi$ plus
the energy stored in $\phi$  yields a Hubble constant of approximately 
$m_{\chi}$. The energy in $\chi$ will decrease during the phase
transition; the energy after the phase transition 
must be very small. In this case the cosmological constant problem is
reduced from $(M_{Pl}/10^{-3}eV)^4\sim 10^{120}$ to  
$(TeV/10^{-3}eV)^4\sim 10^{60}$. For smaller $m_{\chi}$, the reheat
temperature would be lower and the tuning of the cosmological constant
would presumably be smaller.

(II) Energy inflow from extra dimensions. For example, in a
non-elastic scattering of branes, a part of the kinetic energy due
to the relative motion can be converted to radiation on our brane
without changing the brane tension and the cosmological constant. 
For this to work, branes should be sufficiently flat and parallel.

Although both these reheating mechanisms require some form of
fine-tuning, one merit of the mechanism we have proposed is that after
reheating, conventional cosmology with a vanishingly small cosmological
constant would be restored. This is because the coefficient of the
kinetic term is so large at low energy that the scalar field $\phi$
remains frozen  after reheating. Namely, 
%
\begin{equation}
 |\kappa^2\dot{\phi}| \sim
  (\kappa H)^{4m}\times\left|\kappa^2\pi\right|
  \label{eqn:Vdot}
\end{equation}
can be made arbitrarily small at low energy ($\kappa H\ll 1$) by
considering a sufficiently large $m$ since $\pi$ always follows a
regular equation of motion and, thus, evolves continuously. The
stability condition (\ref{eqn:stabilityq=1}) also requires a large
$m$. Since $\phi$ rolls so slowly, our model in general predicts
$w_{\phi}\equiv p_{\phi}/\rho_{\phi}\simeq -1$ today.

Any symmetry restoration or phase transitions that occurs after
reheating would  not pose a problem when $m$ is sufficiently large to
freeze $\phi$. The reason is that the cosmological constant just before 
reheating is the value at zero temperature. The large $m$ ensures that
$\Lambda_{today}$ is still positive and small since $\phi$ continues to
be almost frozen all the way down to the present epoch including the
time when the symmetry is restored and during the time the phase
transition takes place.

If all else fails, although not our initial subjective, the existence of
$\phi$ can at the very least provide a natural framework in which to
implement the anthropic principle. If we assume eternal inflation, there
would be many inflationary universe. In each universe, the cosmological
constant is determined by how much $\phi$ has rolled when inflation
ends. That in turn depends on the number of $e$-foldings that occurred
before inflation stopped. In an eternal inflation scenario, different
numbers of $e$-foldings would occur in different domains and therefore
different $\phi$ values, and hence different values of the cosmological
constant would occur in different regions.


One of the very interesting features of our model is that
despite the singular-looking kinetic term, the scalar dynamics is 
stable for a broad range of parameters. Those parameter choices are (i)
$q>1/2$; (ii) $\alpha>0$; and (iii)  $2(m-1)>q/(2q-1)$. In the
$m\to\infty$ limit, the choice (\ref{eqn:fnear0}) may be replaced by
$f(R)\sim\exp(-\kappa^{-4}R^{-2})$ or similar functions. We would like
to stress again that this condition is imposed only on the most
singular-looking term among many possible terms in the kinetic part and,
thus, adding any kinetic terms which are less singular-looking at $R=0$
does not change anything. Of course, adding a more singular-looking
kinetic term just makes the condition more robust. The more singular a
kinetic term looks, the more stable it is under radiative corrections.

Since $f$ is in the denominator and vanishes in the
$\kappa^2R\to 0$ limit, the kinetic term threatens to be singular at low
energy. Surprisingly, we shall see below that the singular-looking kinetic
term with the large $m$ makes $\phi$ evolve slowly, that the numerator
$K^{q}$ vanishes more quickly than the denominator and that the kinetic
term is actually regular. The more singular a kinetic term looks, the
more regular and stable the dynamics is.

Now let us briefly explain the reasons for the stability
conditions (i)-(iii). (i) For the stability of inhomogeneous
perturbations, it is necessary that the sound velocity squared
$c_s^2=L_{kin,K}/(2KL_{kin,KK}+L_{kin,K})$ is
positive~\cite{Garriga-Mukhanov}. In our model this condition is
reduced to $q>1/2$. (ii) The $R$-dependence of the kinetic term
$L_{kin}$ produces higher derivative corrections to Einstein
equation which might destabilize gravity. As we shall explain
below, the term $\alpha R^2$ can stabilize gravity at low energy
if $\alpha$ is positive. (iii) For the term  $\alpha R^2$ to
control the stability, we need to make sure that the term $\alpha
R^2$ is dominant over $L_{kin}$ at low energy. As shown below,
this is the case if and only if the condition (iii) is satisfied.


We can also show the essential, and somewhat surprising result,
that the standard Friedmann equation is recovered at low energy,
starting from the action (\ref{eqn:action}). There are higher
derivative corrections to the Einstein equation due to the
$R$-dependence of $L_{kin}$ and $\alpha R^2$.

Since there are higher-derivative terms, the stability of the system is
a non-trivial question. In order to see the non-triviality, let us
consider the Klein-Gordon equation $(\Box-M^2)\varphi=0$ as a standard
equation and add $\epsilon(\varphi)\Box^2\varphi/M^2$ to the right hand
side. One might expect that the standard equation should be recovered
whenever $\epsilon\to 0$. Actually, this is not true. The standard
equation is certainly recovered in the $\epsilon\to 0$ limit if
$\epsilon>0$. On the other hand, if $\epsilon<0$ then the system has a
tachyonic degree and is unstable. Moreover, if $\varphi$ crosses a root
of $\epsilon$ then the system experiences a singularity ($\Box^2\varphi$
diverges) and the low energy effective theory cannot be trusted unless a
miracle cancellation occurs.

In our system, specializing to the $q=1$ case again, from the estimate
of $L_{kin}$ in (\ref{eqn:Lkin-is-small}) with (\ref{eqn:stabilityq=1}),
$L_{kin}$ is much smaller than $\alpha R^2$ ($\sim H^4$) at low energy
if $\alpha\ne 0$. Hence, $\alpha R^2$ dominates the higher derivative
corrections and controls the stability. For the theory
$R/2\kappa^2+\alpha R^2$, the parameter $\alpha$ plays the role of
$\epsilon$ above (including the sign) and it is known that the low
energy dynamics is stable if and only if 
$\alpha\ge 0$~\cite{Muller-Schmidt}. Here, stability means that as the
universe expands, the system keeps away from unphysical spurious
solutions and approaches the standard low energy evolution
asymptotically. If we did not include the term $\alpha R^2$ then
$L_{kin}$ would make the quantity corresponding to $\epsilon$ above to
be negative essentially because $f(R)$ is in the denominator. Therefore
our system is stable and the standard Friedmann equation is recovered at
low energy if and only if $\alpha>0$ and (\ref{eqn:stabilityq=1})
(the condition (iii) above for a general $q>1/2$) are satisfied.


A possible source of instability which would disturb weak gravity in a
Minkowski background could be an inhomogeneous fluctuation of $\phi$,
which could possibly generate violent breakdown of linearized Einstein
gravity. Quite surprisingly, this is not the case and the seemingly 
most dangerous part, the kinetic term, is not as dangerous as it
looks. Essential to this conclusion is the constraint equation, which
prevents $\phi$ from fluctuating freely and forces the denominator and
the numerator to fluctuate in a strongly correlated way so that the
contributions of the kinetic term to the equation of motion are regular
and much smaller than those of the $\alpha R^2$
term~\cite{paper2}. Since contributions of the singular-looking kinetic
term are small enough, the only possibly important correction to
Einstein gravity is again due to the $\alpha R^2$ term. This tells us
that the linearized gravity in our model should be identical to that in
the theory $R/2\kappa^2+\alpha R^2$. Hence, we should be able to recover
the linearized Einstein gravity in Minkowski background at distances
longer than the length scale $l_*=\sqrt{\alpha}\kappa$ and at energies
lower than $l_*^{-1}$~\cite{Stelle}.


Before ending this letter, it is perhaps worth while stressing
again that the way we have achieved the vanishing cosmological
constant is stable under radiative corrections. Note that our
essential assumption is that at least one coefficient in the
kinetic part diverges at zero curvature. Although we did not give a
reason why the divergence occurs at zero curvature, we showed that this
assumption is stable under radiative corrections and leads to
interesting dynamics. Adding extra terms to the action, irrespective of
whether they are in the kinetic part or in the potential part, never
spoils this assumption and, thus, the mechanism is stable under
radiative corrections. The more singular a kinetic term looks, the more
stable it is under radiative corrections.

Even more surprisingly, the seemingly most dangerous kinetic term
does not lead to unstable dynamics.  In fact, the more singular a
kinetic term appears, the more regular and stable the dynamics is.
In particular, we have shown not only that the cosmological constant
vanishes but also that the standard Friedmann equation and the
linearized Einstein gravity are recovered at low energy.  This
point is in itself of interest since it means there are new types
of dynamics that can lead to the same type of weak gravity on a
Minkowski background that we see today.

The feedback mechanism we have proposed can be considered as a way
to protect a zero or small cosmological constant against radiative 
corrections. Hence, despite the necessity for late reheating and the
lack of our knowledge about a parent theory, this may be the right
starting point for thinking about the cosmological constant.

\begin{acknowledgments}
 We would like to thank N.~Arkani-Hamed, J.~Bloom, R.~Bousso,
 H.~C.~Cheng, P.~Creminelli, E.~Flanagan, L.~Motl and A.~Strominger for
 useful comments. SM's work is supported by JSPS. This work was
 supported in part by NSF grant PHY-0201124 and PHY-9802709.
\end{acknowledgments}


\end{document}